\documentclass{article}
\usepackage{spconf,amsmath,epsfig,graphicx}

\begin{document}\sloppy

\def\x{{\mathbf x}}
\def\L{{\cal L}}

\title{CNN BASED MUSIC EMOTION CLASSIFICATION}
%
\name{Xin Liu, Qingcai Chen*, Xiangping Wu, Yan Liu, Yang Liu}
\address{Harbin Institute of Technology, Shenzhen, China\and Department of Computing, The Hong Kong Polytechnic University, Hong Kong\and Department of Computer Science, Hong Kong Baptist University, Hong Kong
\and Email:\{hit.liuxin, qingcai.chen, wxpleduole\}@gmail.com, csyliu@comp.polyu.edu.hk, \and csygliu@comp.hkbu.edu.hk}

\maketitle

\begin{abstract}
Music emotion recognition (MER) is usually regarded as a multi-label tagging task, and each segment of music can inspire specific emotion tags. Most researchers extract acoustic features from music and explore the relations between these features and their corresponding emotion tags. Considering the inconsistency of emotions inspired by the same music segment for human beings, seeking for the key acoustic features that really affect on emotions is really a challenging task. In this paper, we propose a novel MER method by using deep convolutional neural network (CNN) on the music spectrograms that contains both the original time and frequency domain information. By the proposed method, no additional effort on extracting specific features required, which is left to the training procedure of the CNN model. Experiments are conducted on the standard CAL500 and CAL500exp dataset. Results show that, for both datasets, the proposed method outperforms state-of-the-art methods.
\end{abstract}
\section{Introduction}

It is well known that different type of music make quit different influences on our emotion. Researchers have shown that, music, explained as ¡°organized sound¡±, can resonate with our nerve tissue~\cite{liu2015what}. As researchers mentioned, for little babies who still do not know what music is, what language is and even what they see is, can make responds to what they hear. It is more likely a biological instinct, just the interaction between sound rhythm or melody and their brain~\cite{n1998m,Sloboda01}. In fact, the delicate relationship between music and emotion has already been explored by numerous researchers for a long period~\cite{j2010e}. Barthet et al.~\cite{b2012m} gave a detail introduction about music emotion recognition (MER) task. Wieczorkowska et al.~\cite{Wi} firstly formulated MER as a multi-label classification problem. And then many researchers follow this opinion. In multi-label classification, each training or test music segment sample is a sequence of features that have been assigned multiple labels. Each label indicates one type of emotion.

For a classification task, feature selection is one of the most important tasks. Though there is still no standard guidance for selecting features that contribute most to the representation of music~\cite{s2011a}, acoustic features are still most prevalent in the feature selecting procedure~\cite{e2015e}. Acoustic features mainly consist of rhythmic features, timbre features and spectral features~\cite{M2014MEC}. Rhythmic features are derived by extracting periodic changes from a beat histogram~\cite{M2014MEC}. Timbre features consist of a series of Zero Crossing Rate, MFCC~\cite{l2000m} and Chroma~\cite{s2010f}. Spectral features include Spectral Flatness Measure, Spectral Centroid, Spectral Crest Factor, Spectral Rolloff and Spectral Flux. All these features can represent music respectively or mutually in a numeric way. In addition to features selection, the other important task of MER is the choosing of classifier. Researchers have tried some classifiers. For example, Calibrated Label Ranking classifier using a Support Vector Machine (CLRSVM)~\cite{f2008m}, Random k-Labelsets (RAkEL)~\cite{t2011r}, Back-propagation for Multi-Label Learning (BPMLL), Multi-Label K-Nearest Neighbor (MLkNN)~\cite{z2007m} and Binary Relevance kNN (BRkNN) etc. Among these classifiers, in most cases, CLRSVM outperforms the rest~\cite{b2012m}.

Though great improvements have been made by researchers, the state-of-the-art of MER is still far from satisfactory. Considering the inconsistency of emotions inspired by the same music segment for human beings, seeking for the key acoustic features that actually affect on emotions is a really challenging task and is a crisis obstacle of solving MER problem. To address this problem, in this work, we propose a novel model based on deep CNN architecture. This model directly uses the spectrogram of music audio without complex artificially selected features. Our main contributions include: 1) a novel CNN based MER model that only uses the music audio spectrogram as input is proposed. By this model, manually selection of complicate features is avoided, which not only simplifies the process of model construction, but also keeps most of the original time and frequency domain information. 2) The convolution method on local time and frequency of spectrogram is proposed to address the issue of variance in time length for different music segments. 3) MER experiments are conducted on standard CAL500 and CAL500exp dataset and the results show that the proposed method outperforms existing methods and gets obvious improvements on state-of-the-art F1 measure.

This work is organized as follows. In Section 2 we detail describe the spectrogram for specific data and the structures of our deep neural network. In Section 3 we discuss the experiments on the dataset CAL500~\cite{t2007t} and CAL500exp~\cite{w2014t} and compare the results with state-of-the-arts algorithms. Finally, some conclusions and future work are drawn in Section 4.
\begin{figure*}
\begin{center}
 \includegraphics[width=1\textwidth]{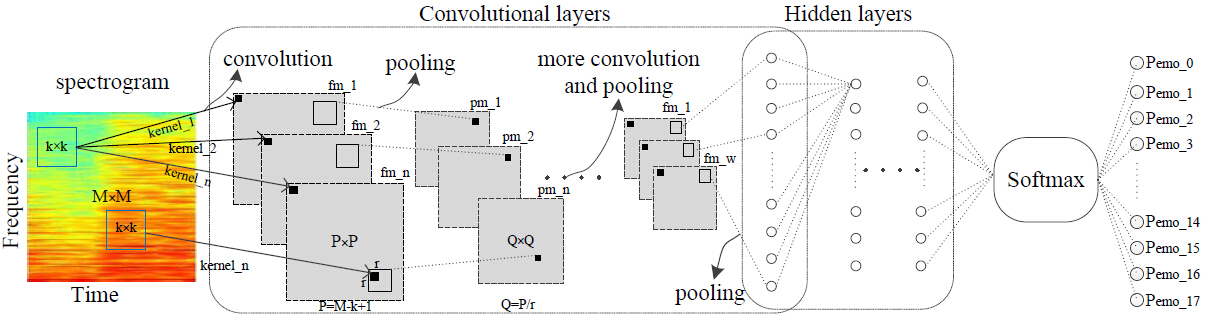}
 \caption{Flowchart of CNN based music emotion recognition. Start from original music files, then go through convolutional layers and hidden layers, and make predictions with a classifier SOFTMAX at the end.}
\end{center}
\end{figure*}
\section{PROPOSED METHOD}

\subsection{CNN framework for music emotion classification}
Fig.1 shows the CNN framework for the music emotion classification. Unique from existing methods, this framework uses only the spectrogram of the audio signal as input. And a network with few convolution-pooling layers, hidden layer and a SOFTMAX classifier is then constructed to extract features and classify the emotions of given music segment.
As well known that emotions inspired by a segment of music are close relevant to both the rhythm and melody of music~\cite{m2012e}, which are mainly determined by the distribution and variance of signal energy on time and frequency domains. Spectrogram, as a type of representation for variances of frequency spectrum with time, not only presents a visualization tool, but also an important type of rich-information feature for audio signal analysis~\cite{c2011m,F2007S}. The spectrogram is computed via the Short-Time-Fourier-Transformation of audio signal along the time axis as below formula~\cite{Hebbal2014fpga}:
\begin{equation}
Spec(t,f)=|nfft(t,f)|^{2},
\end{equation}
Though the phase information is lost by transform audio signals into their spectrogram, the power variance information along both frequency and time axis are well presented~\cite{F2007S}.  As an example, Fig. 2 shows the waveforms and spectrograms of two music segments with emotion tags of ¡°emotional, exciting, happy and powerful¡± and ¡°calming, tender, mellow and pleasant¡± respectively.  It is obvious that from Fig. 2 we can more easily tell the differences between the two music segments via the spectrograms than though their waveforms.

Considering the capabilities shown by CNN on the capturing of complexity features for 2 or 3-dimensional pictures, using CNN to extract features from spectrogram may also be an effective way. To verify it, in this paper, the general CNN layers for feature extraction of music spectrogram are introduced. As shown in Fig. 1, each spectrogram is convolved with n convolution kernel functions and n feature map matrices are produced correspondingly. The pooling operations are then executed on each feature map with multiple kernels to generate self-adapting features by neural network. More convolution and pooling operations are further conducted according to the application requirements.

In fact, CNN operations executed on spectrogram are independent from the emotion classification and thus applicable in any music processing tasks. For the goal of this paper, a multi-layer perceptions (MLP) with hidden layers and softmax operations for specific emotion tag set are presented as the last part of the framework. The input layer of the MLP is generated by stretching and concatenating the matrixes of the last CNN layer. To tuning the whole CNN based network, supervised learning is conducted by the differences of the softmax outputs with standard emotion tags of the input music segment.

In following sections, we detail describe the main processing of different stages in above framework.
\begin{figure}
\begin{minipage}[b]{1.0\linewidth}
  \centering
  \includegraphics[height=2.5cm,width=1\linewidth]{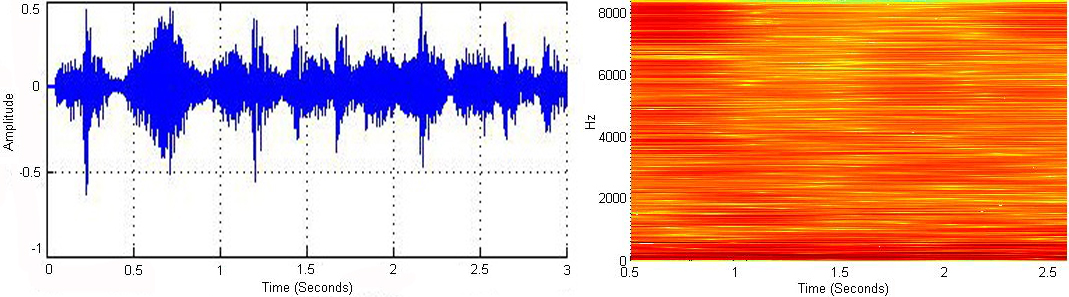}
  \includegraphics[height=2.5cm,width=1\linewidth]{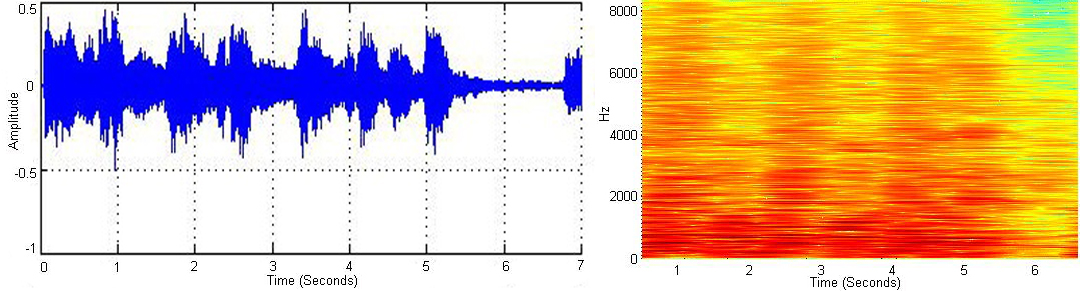}
  \caption{Waveform and spectrogram of two segments with emotion tags of similar semantics between two different music. a) emotional, exciting, happy and powerful. b) calming, tender, mellow and pleasant.}
\end{minipage}
\end{figure}
\subsection{Preprocessing of music spectrogram}
The preprocessing of spectrogram is a key point of successfully applying CNN on music spectrogram. It is because that the input to CNN is required as a fix dimensional matrix, while the length of music segment given for emotion classification is usually variant duration.

As Eq. (1) shows that, the spectrogram of a music segment is produced by computing a series value sets of the discrete-time \(STFT(t, f)\) at different time points of the music segment along the time axis. At each time point, the number of values produced along frequency axis is determined by the needs of frequency resolution for a given application. For this reason, we firstly determine the frequency points \(K\) of the \(STFT(t, f)\). Then the number of time points is set as the same of the frequency points to generate a \(K\times K\) matrix. Since we directly using matlab to generate the spectrogram, in this paper, we select the equal-distance time point by computing the overlap signal number of two concatenated windows for a given music segment as below:
\begin{equation}
N_{overlap}= \frac{(M+1)*N_{win\_szie}-N_{music\_len}}{M}\
\end{equation}
Here, \(M\), \(N_{win\_szie}\) and \(N_{music\_len}\) denote the frequency dimension, window length and number of signals contained in the given music segment. Negative overlap signal means number of skipping signal from previous window to the next window. To simplify the process, the window length is equal to \(nfft\), the number of points for the FFT. The frequency dimension \(M=nfft/2+1\).

To construct the fixed dimensions of input matrix for CNN network, there must be more sophisticate methods existing. But our goal is focused on the effectiveness of CNN architecture for spectrogram based music emotion classification, so which is left for further study.
\subsection{Convolutional neural network for music spectrogram}
As mentioned in the previous section, the emotions formed when people listen to music mostly will go through a cumulative process as time goes on. Then this is another reason why we choose spectrogram to represent one music. In convolutional neural network, the input feature maps will be calculated through the operation called local field to generate new feature maps. The concrete implementation in the network is that the nodes for calculating in the same layers are neighboring and the nodes in two adjective layers are not fully connected. So this operation can satisfy the need of calculating the relevance of adjacent time and frequency.

\begin{figure}
\begin{minipage}[b]{1.0\linewidth}
  \centering
  \includegraphics[width=1\linewidth]{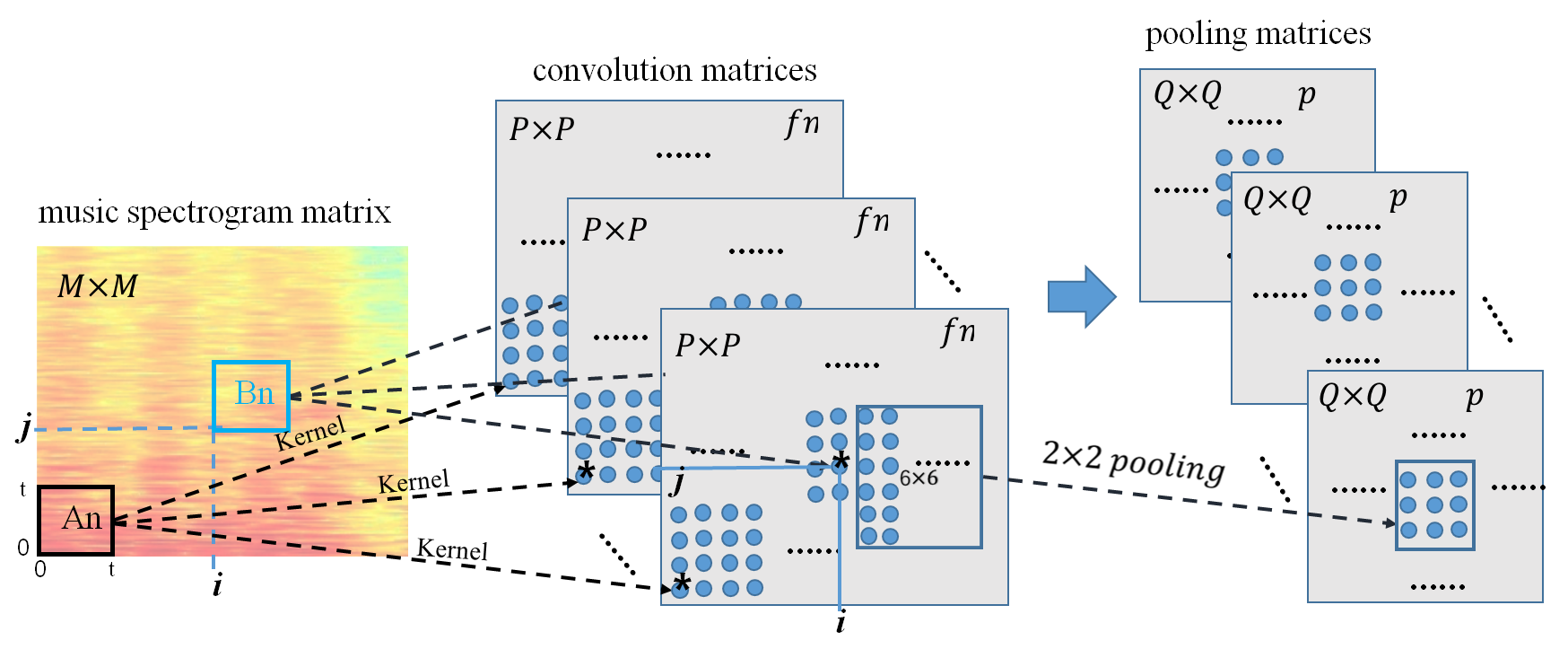}
  \caption{Convolution operation on local time and frequency of spectrogram. The left color picture represents the input music spectrogram. The middle white squares are new feature maps after convolution operations. The right white squares are new pooling maps after pooling operations.}
\end{minipage}
\end{figure}
As shown in Fig.3, it shows the local convolution operation. In the stage of spectrogram, there are a lot of windows like area A and B. The values in one area are calculated by some filters through convolution computing, each filter will generate one feature map. The convolution calculation~\cite{B95c} is
\begin{equation}
z_{i}^{l}=\sigma(w^{(l,f_k)}z_{j*}^{l-1}+b^{(l,f_k)})
\end{equation}
where \(z_{i}^{l}\) means the output of feature map in layer \(l\) at location \(i\) and corresponds to the one point of \(A_n\) or \(B_n\) in Fig.3, \(w^{(l,f_k)}\) is the parameters of k-th filter from layer \(l-1\) to \(l\) and \(f_k\) is some filter in \(F_n=[f_0,f_1,...,f_n]\), \(b^{(l,f_k)}\) is the bias for layer \(l\), \(z_{j*}^{l-1}\) means the areas of location \(j\) in layer \(l-1\) and corresponds to the area \(A\) or \(B\) in Fig.3, \(\sigma(.)\)is the activation function (e.g., Sigmoid or Relu~\cite{d2013im}).

For example, area \(A\) is a square of time varying from \(t0\) to \(t1\) and frequency varying from \(f0\) to \(f0\). Through different filters, we will get the subarea \(A_1\) \(A_2\) to \(A_n\) in next feature maps. Area \(B\) is the same as \(A\). Other convolutions also act as these steps. This is also simulating the procedure of generating emotions over time and frequency. After each convolution, we take a max-pooling in every unit window for every filter as

\begin{equation}
z_{i}^{l,f_k}=max(z_{2i-1}^{l-1,f_k},z_{2i}^{l-1,f_k})
\end{equation}
this means the output of pooling map is the maximum value selected from the pooling window of its feature map. In Fig.3, after max-pooling, we will get such output \(a_n\) or \(b_n\) in pooling maps.
In convolutional layers, input feature map is calculated through repeated convolution-pooling operation. The difference among them is that the different filters for generating new feature map. In the final step, the output size of feature maps is a vector, so we just need to reshape these nodes as the final output features of convolutional layers for hidden layer.
\subsection{Emotion classification based on CNN of music spectrogram}
In hidden layer and classifier layer, the input is the same as the reshaped vectors in previous layer. The nodes in hidden layer are fully connected. Then after a few full-connected layers, we may get fixed vector as the input of softmax to classify. For softmax, the dimension of output equals to the number of emotion tags. Each dimension corresponds to one tag. Once the numerical value exceeds a certain threshold, we will conclude that the tag belongs to this music. We also train our neural network in some valid techniques, such as dropout~\cite{d2013im}, activation function and so on.

\section{EXPERIMENTS}

\subsection{Dataset}
We tested the music emotion recognition performance of the proposed approach. A series of experiments were performed on CAL500exp and CAL500. CAL500exp is an enriched version of the well-known CAL500. Wang et al.~\cite{w2014t} published the dataset. Labels of CAL500exp are annotated in the segment level instead of track level in CAL500. In the other words, in CAL500exp, each song contains several segments split from itself. And each segment is annotated as a dependent data from 18 emotion tags. So in CAL500exp dataset, there are total 3223 items. While in CAL500, each whole song is regarded as one train or test data and each song is scored on 18 emotions form 1 to 5 by different listeners, then we confirm the emotion labels of this song by the means that at least 80 percent of all listeners agree the score~\cite{t2007t}. The number in CAL500 is quite less than CAL500exp. There are about 502 items.

\subsection{Evaluation criterion}
As a multi-label task, some common criteria were used to evaluate the performance, namely label-based metrics and  example-based metrics. Most researchers use macro average and micro average to evaluate the overall performance across multiple labels.

In this paper, we will focus on the criteria of precision (P), recall (R) and F1 score which considers both the precision and the recall. As a contrast, some other criteria such as hamming loss, AUC score, average precision score(AP) and one error will also be introduced in brief. The computational method can be found from python-sklearn which is a python module for machine learning and data mining at~\cite{p2011s}. For some criteria including F1 score, AUC score and average precision, the larger the metric value is, the better performance it shows. As for another two metrics hamming loss and one error, it is just the opposite. Smaller values indicate the better performance. For each train set, we stop the training procedure until CNN reached certain epoch iteration.

\subsection{Cross validation}

we constructed train set, validation set and test set based on these segments and whole songs. We proposed ten-fold cross validation for train, validation and test. In both data set, we performed ten-fold cross validation randomly. 3223 segments and 502 songs were first disrupted into a random order, then we selected 10 percent of the same unordered set as test set and the rest as train or validation set for ten times on each dataset respectively.

Ten-fold cross validation aims to make a contrast with those state-of-the-arts already known. In CAL500exp, each fold contains the similar number of train and test set with the average differences in 5, namely about 2902 and 321. While in CAL500, each fold contains 452 songs for train and 50 for test. Since CNN model generates a series of parameters in each epoch iteration and each will bring different results. So parts of both train set are used for validation to find proper parameters. In our experiments we choose ten epochs as benchmark based on validation set. And the average value of ten epochs represents the performance of this fold.

\subsection{Model structure}

As shown in Fig. 1, the initial input is the spectrogram with fixed size. Compared to the general input size of CNN, our input structure may be a little complex. But such a scale can retain more information of a long music. In price, more calculations and memory space will be cost. Besides we need to know that the size is not the more complex the better. If more complex, there will be more redundant information which will lead to a negative impact on the final result. In order to balance the loss of information and the cost of time and space complexity, we need to select a balanced size. In our model, there are two key points that we need to plan. One is the size of input spectrogram, the other is the structure of CNN.

First we need to know is that the size of spectrogram is associated with the length of music. After analyzing music in both dataset, we find that one song in CAL500 is about 5 minutes long while 5 seconds more or less for one segment in CAL500exp. So some test experiments are conducted to determine the size of spectrogram. In most case, we will make it square matrix. Since the vertical dimension is calculated based on nffts by half of nffts plus one, we may try different sizes by changing nffts to 256, 512, 1024 and other else. We set the initial size as \(257\times 257\) by rule of thumb, so all test sizes are \(1025\times 1025\), \(513\times 513\) and \(129\times 129\) in total. We take final performance and time cost into consideration for each input size in a same structure.

\begin{table}[t]
\begin{center}
\caption{Experiments on different size of spectrogram. Each line represents one experiment with different size. And middle two columns are macro f1 and micro f1 respectively. The last column is the time cost when training the model for 800 epoch.} \label{tab:cap}
\begin{tabular}{|c|c|c|c|}
  \hline
  Experiment & macro\_f1 & micro\_f1 & time(hour)
  \\
  \hline
  Size129 &	0.391 &	0.466 &	15  \\
  Size257 &	0.407 &	0.463 &	55  \\
  Size513 &	0.410 &	0.484 &	170  \\
  Size1025 &	0.412 &	0.487 &	400  \\
  \hline
\end{tabular}
\end{center}
\end{table}
Table 1 shows these experiments, we find that if we regard \(257\times 257\) as baseline at both considerations, bigger size such as \(1025\times 1025\) and \(513\times 513\) can get better performance and smaller size shows poor performance, but \(1025\times 1025\) take much more time cost that is far beyond the other size. With overall consideration, the spectrogram in CAL500 is \(513\times 513\) and in CAL500exp it is \(257\times 257\).

Next one we need to know is the structure of CNN including total layers and nodes in each layer. We first define two kinds of neural network, one simple network contains four convolutional layers with nodes no more than 50 in each and one hidden layer, the other one seems more complex with nodes from 100 to 200 in each and three or more hidden layers.  The simple contrasts on macro metric of both dataset on different network are given in Table 2.
\begin{table}[t]
\begin{center}
\caption{Contrast of different networks on both dataset. Symbol ¡°Sim¡± and ¡°Com¡± mean simple and complex network respectively. CAL500exp is short for symbol ¡°exp¡± and CAL500 for ¡°500¡±.The combinations of different symbols represent the experiment of the dataset on this kind of network.} \label{tab:cap}
\begin{tabular}{|c|c|c|c|}
  \hline
  Experiment & P & R & F
  \\
  \hline
  Sim\_exp &	0.603 &	0.614 &	0.596 \\
  Com\_exp &	0.614 &	0.575 &	0.583 \\
  Sim\_500 &	0.426 &	0.583 &	0.472 \\
  Com\_500 &	0.418 &	0.513 &	0.437 \\
  \hline
\end{tabular}
\end{center}
\end{table}
From Table 2, we can see that simple network outperforms complex network nearly on all metrics for both
dataset. Then through these experiments, we confirm the size of spectrogram and structure of CNN.
\subsection{Results}

In previous parts, we introduce the size of spectrogram and structure of CNN. During training our model, we need to adjust the parameters in the network. We train the model with iterations(also called epoch). And in each iteration, we will get a cost to indicate the processing of current epoch. The cost can not only tell us whether the model is working but also help us to select the parameters for testing.
\begin{figure}
\begin{minipage}[b]{1.0\linewidth}
  \centering
  \includegraphics[width=1\linewidth]{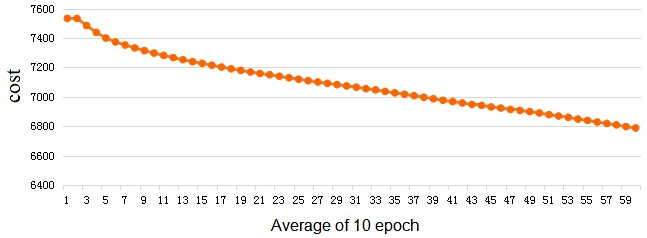}
  \caption{Cost reduction with the increasing of iterations. The serial number in horizontal coordinate is the average of 10 epoch, and longitudinal to their average cost.}
\end{minipage}
\end{figure}

Fig.4 gives an example of cost reduction with the increasing of iterations when training the model. Since there may be hundreds of epochs, we choose the average cost of ten epochs as one coordinate. The trend of cost indicates the validity of model and allows us to do test on different epochs to find the optimal parameters.

After a series of certain iterations, we finish the train. According to the validation set and parameter adjustment,  we obtain the performance of all metrics on both dataset. The final results of ten-fold cross validation with our model are shown in Table 3. Since we cannot try all parameters that represent convergence, there will be max 0.002 errors for the result. Wang et al.~\cite{w2014t} report the results when they published the CAL500exp dataset. The results are as shown in Table 3. And on CAL500, the results are shown in Table 4, compared with~\cite{w2014t} and~\cite{liu2015what}.

From Table 3 and 4, we can see that through the contrast of those results published, our proposed model outperforms the state-of-the-art on the metrics of macro F1 and micro F1.

\begin{table}[t]
\begin{center}
\caption{Results on CAL500exp published. The first line of numerical parts shows the result of~\cite{w2014t}. We list both of our results at the following two lines on CAL500exp.} \label{tab:cap}
\begin{tabular}{|c|c|c|c|}
  \hline
  Description & P & R & F
  \\
  \hline
 ~\cite{w2014t} Macro average &	0.455 &	0.759 &	0.561 \\
  CNN Macro average &	0.603 &	0.614 &	0.596 \\
  CNN Micro average &	0.686 &	0.735 &	0.709 \\
  \hline
\end{tabular}
\end{center}
\end{table}

\begin{table}[t]
\begin{center}
\caption{Results on CAL500 published. The first line of numerical parts shows the result of~\cite{w2014t}. The second and third line of numerical parts shows the result of~\cite{liu2015what}. The final two lines are our results.} \label{tab:cap}
\begin{tabular}{|c|c|c|c|}
  \hline
  Description & P & R & F
  \\
  \hline
  ~\cite{w2014t} &	0.301 &	0.701 &	0.417 \\
  Macro average~\cite{liu2015what} &	0.438 &	- &	0.444 \\
  Micro average~\cite{liu2015what} &	0.476 &	- &	0.395 \\
  CNN Macro average &	0.426 &	0.583 &	0.472 \\
  CNN Micro average &	0.459 &	0.640 &	0.534 \\
  \hline
\end{tabular}
\end{center}
\end{table}

\begin{table}[t]
\begin{center}
\caption{Results of hamming loss, AUC score, average precision and one error on CAL500 and CAL500exp.} \label{tab:cap}
\begin{tabular}{|c|c|c|c|c|}
  \hline
  Dataset &	Hamloss &	AUC &	AP &	One-error
  \\
  \hline
  CAL500CNN &	0.325 &	0.675 &	0.458 &	0.423 \\
  CAL500~\cite{liu2015what} &	0.193 &	- &	0.803 &	0.250 \\
  500expCNN &	0.212 &	0.799 &	0.629 &	0.120 \\
  500expe{w2014t} &	- &	0.884 &	- &	- \\

  \hline
\end{tabular}
\end{center}
\end{table}

Table 5 gives the results of the rest metrics on CAL500 and CAL500exp. From Table 5, we can also see that the variable trend of the rest metrics is also in accordance to F1 score on both dataset. But the D-value differs greatly in each metric between CAL500 and CAL500exp. We guess that the reason mainly lies in the following two aspects. One point is the data quantity, especially for CAL500. After the whole set was divided into validation and test set in the proportion of 10\%, the number for train was greatly reduced which  actually is inadequate for deep neural network. The other one is that the criterion to ascertain labels for music is ambiguous in CAL500. Not like explicit labels of each music in CAL500exp, there is only a series of scores between 1 and 5 on each label from several different users. Then we need to stipulate labels in some way. This may cause inaccurate compared to CAL500exp. Both aspects may be the exact cause of the results above.

\section{CONCLUSIONS AND DISCUSSIONS}

In this work, we propose a novel method combining original music spectrogram with deep convolutional neural network (CNN) to predict the emotion tags.
Though we have got a obvious performance gain by this model, there are still a lot works to do since the final result of the method is only 0.709 on micro F1 measure. First, this CNN architecture is not fine-tuned, there is still a lot of improved space on both the convolution step and the feedback training step. Second, for a song or a segment of song with tagged emotions, only fixed numbers of segments are sampled and there is no detail study on the selection of time or frequency points. Though the advantages of CNN on extracting useful features from raw data, there is still no research on the meaning of CNN outputs, which makes it difficult to further understand and deduce the source of inspired emotions for a given music.

\bibliographystyle{IEEEbib}
\bibliography{icme2016template}

\end{document}